**Key Points:**
- PCA of Birkeland currents shows regions 1 and 2 pattern most prevalent
- Cusp currents are strongest in summer when ionospheric conductance is high
- There is no one pattern that appears consistently with substorms

**Correspondence to:**
S. E. Milan,
steve.milan@le.ac.uk

# Principal component analysis of Birkeland currents determined by the Active Magnetosphere and Planetary Electrodynamics Response Experiment

**S. E. Milan[1,2], J. A. Carter[1], H. Korth[3], and B. J. Anderson[3]**

[1]Department of Physics and Astronomy, University of Leicester, Leicester, UK, [2]Birkeland Centre for Space Science, University of Bergen, Bergen, Norway, [3]Johns Hopkins University Applied Physics Laboratory, Laurel, Maryland, USA

**Abstract** Principal component analysis is performed on Birkeland or field-aligned current (FAC) measurements from the Active Magnetosphere and Planetary Electrodynamics Response Experiment. Principal component analysis (PCA) identifies the patterns in the FACs that respond coherently to different aspects of geomagnetic activity. The regions 1 and 2 current system is shown to be the most reproducible feature of the currents, followed by cusp currents associated with magnetic tension forces on newly reconnected field lines. The cusp currents are strongly modulated by season, indicating that their strength is regulated by the ionospheric conductance at the foot of the field lines. PCA does not identify a pattern that is clearly characteristic of a substorm current wedge. Rather, a superposed epoch analysis of the currents associated with substorms demonstrates that there is not a single mode of response, but a complicated and subtle mixture of different patterns.

## 1. Introduction

The Active Magnetosphere and Planetary Electrodynamics Response Experiment (AMPERE) [*Anderson et al.*, 2000; *Waters et al.*, 2001; *Anderson et al.*, 2002, 2008] has provided measurements of the Birkeland currents or field-aligned currents (FACs) in the Northern and Southern Hemispheres at 10 min cadence from 2010 to 2013, using magnetometer observations from the Iridium constellation of close to 70 satellites. A number of studies have employed AMPERE observations to study the structure of the current systems and their response to solar wind-magnetosphere coupling. For instance, *Clausen et al.* [2012, 2013a, 2013b] and *Coxon et al.* [2014a, 2014b] have demonstrated that the region 1 and 2 current systems [*Iijima and Potemra*, 1976] observed by AMPERE undergo cycles of expansion to lower latitudes and contractions to higher latitudes in tune with the substorm cycle, and *Anderson et al.* [2014] showed dayside followed by nightside activations of the current systems in response to southward turnings of the interplanetary magnetic field (IMF). Both sets of observations are consistent with the expanding/contracting polar cap (ECPC) model of the Dungey cycle of magnetospheric and ionospheric convection [*Dungey*, 1961; *Siscoe and Huang*, 1985; *Cowley and Lockwood*, 1992; *Lockwood and Cowley*, 1992; *Milan et al.*, 2007; *Milan*, 2015]. In addition, *Murphy et al.* [2013] and *Sergeev et al.* [2014] have used AMPERE observations to explore the structure of nightside FACs during substorms, specifically attempting to elucidate the structure of the substorm current wedge [*McPherron et al.*, 1973], while *Wilder et al.* [2013] studied cusp currents associated with magnetic tension forces during periods of nonzero IMF $B_Y$.

The aim of the current paper is to use principal component analysis [*Jolliffe*, 2002] to determine the repeatable patterns of currents that make up the constantly varying observations. Principal component analysis (PCA) decomposes a data set into a series of basis functions that reveal the structure of the underlying correlations within the data. For instance, PCA has also been used for facial recognition [*Sirovich and Kirby*, 1987; *Turk and Pentland*, 1991] and it is potential for recognizing otherwise hidden structure within AMPERE current maps that motivates our study. Recently, PCA has seen wide adoption by the space and geophysics community [e.g., *Ohme et al.*, 2013; *Natali and Meza*, 2010]. *He et al.* [2012] used an empirical orthogonal function decomposition (closely related to PCA) to investigate FAC patterns observed by the CHAMP satellite. *Kim et al.* [2012] applied PCA to polar cap ionospheric convection observations and found two dominant modes of response: uniform antisunward flow, correlated with the $B_Z$ component of the IMF, and a mode which introduced a dawn-dusk asymmetry in the flows, correlated with IMF $B_Y$. *Cousins et al.* [2015] performed PCA on AMPERE current maps and found two dominant components related to IMF $B_Z$ and $B_Y$, the FAC counterparts of the convection modes of *Kim et al.* [2012], and a third related to expansions and contractions of the polar cap.

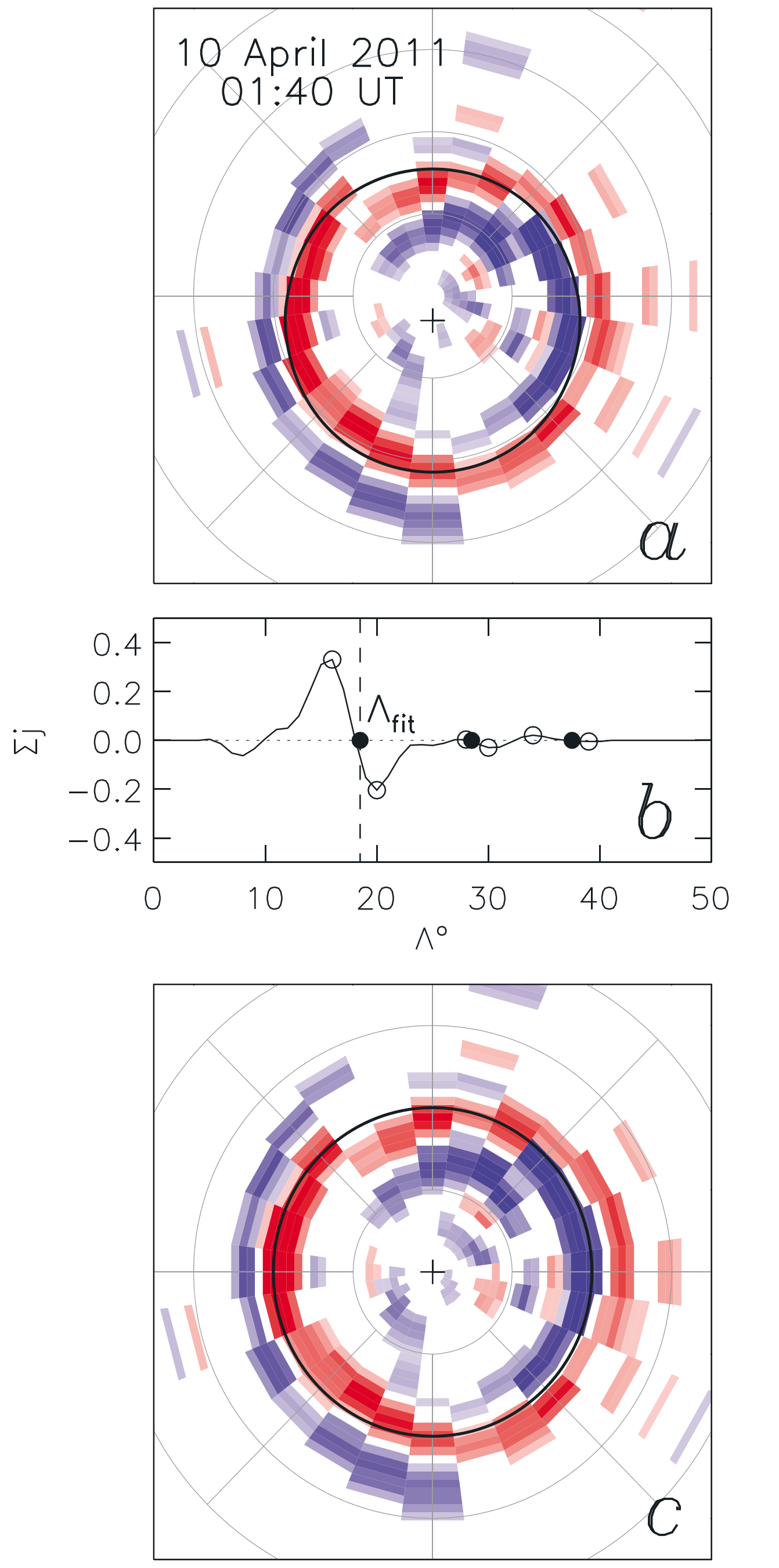

**Figure 1.** (a) An example AMPERE current map from 01:40 UT, 10 April 2011, on a magnetic latitude and local time grid, with 12 magnetic local time (MLT) at the top. Red and blue indicate upward and downward FACs, between ±1 μA m$^{-2}$, respectively. The cross displaced toward midnight from the geomagnetic pole is the assumed centroid of the region 1/2 current system. The black circle of radius $\Lambda_{fit}$ is fitted (see below) for normalization purposes. (b) The summed currents around the circumferences of circles of radii $\Lambda$, centered on the cross in Figure 1a. Positive/negative bipolar signatures are identified (open circles), and the zero crossings (closed circles) are found. The largest bipolar signature is identified with the R1/R2 current system, and the radius of the corresponding zero crossing, $\Lambda_{fit}$, is used for normalization. (c) The normalized current map, centered on the geomagnetic pole and stretched to a normalization radius of 20°, is indicated by the black circle.

In this paper we perform PCA on AMPERE current maps that have been preprocessed to remove variations in the radius and center of the current ovals (related to the ECPC), so as to reduce smearing and to better resolve small-scale features in the patterns. We then investigate the response of the principal components to solar wind parameters and the occurrence of substorms.

## 2. Observations and Discussion

This study employs AMPERE observations of the Birkeland currents from the Northern Hemisphere only. Each AMPERE map covers the region poleward of 50° geomagnetic colatitude, on a 24 × 50 grid in magnetic local time and colatitude, centered on the northern or southern geomagnetic pole. Although AMPERE provides current maps with 2 min cadence, these are produced by a sliding 10 min average of the Iridium observations, so 144 independent maps are available per day. An example map is shown in Figure 1a. The region 1 and 2 (R1/R2) current system is clearly visible as upward/downward (red/blue) currents near a latitude of 70°. The polarity of the currents is opposite in the dawn and dusk sectors.

The size of the FAC pattern changes continuously due to expansions and contractions of the polar cap [*Milan et al.*, 2003, 2007; *Clausen et al.*, 2012]. Before undertaking PCA, we normalize the current patterns to a consistent size to remove this effect. Using a technique similar to that described in *Milan* [2009], we assume that the R1/R2 current regions are approximately circles centered on a point displaced a few degrees antisunward from the geomagnetic pole. For instance, in Figure 1a it is assumed that the center of the current systems (marked by a cross) is offset by 3° of latitude, $\Lambda_0$, along the midnight meridian.

We determine the current strength integrated around circles of differing radii, centered on this point. Consider a circle of radius $\Lambda$. We find the mean of the current density, $j$, at 48 equally spaced points around the circumference of the circle, first multiplying currents in the dusk sector (12–24 MLT) by −1. In this way, if the circle coincides with region 1 currents, a positive value is measured, and a negative value is measured if the circle coincides with region 2 currents. The variation of this sum, $\Sigma j$, with $\Lambda$ is shown in Figure 1b, clearly showing a bipolar signature associated with the R1/R2 currents.

The largest bipolar signature in $\Sigma j$ is identified, and the latitude of the zero crossing, $\Lambda_{fit}$, is used as the boundary between R1 and R2 currents. This procedure is repeated with values of $\Lambda_0$ between 0° and 5°, and the combination of $\Lambda_0$ and $\Lambda_{\rm fit}$ which gives the greatest R1/R2 peak-to-peak variation is used as the fit. This is indicated in Figure 1a by the black circle. If the currents are too weak, such that the bipolar signature represents an average peak-to-peak current density of less than 0.15 μA m$^{-1}$, then the fit is considered unreliable and the map is discarded. Seventy-nine percent of maps were successfully fitted.

The current map is then stretched onto a 24×40 grid (with $n = 960$ elements), such that the reference circle is centered on the pole and has a radius of 20°. The resulting pattern is shown in Figure 1c. In all, 123,340 maps ($m = 123,340$) from 2010 to 2012 were normalized in this fashion.

We performed principal component analysis on these normalized maps. Each map is represented by an $n$−dimensional vector, **J**. Each **J** is mean centered—that is, the mean of all the elements, $\frac{1}{n}\sum_{i=1}^{n} J_i$, is subtracted from each element—but otherwise no scaling is applied. The $n \times m$ matrix **X** represents all the observations, being a concatenation of all the **J**s. From this, the covariance matrix $\boldsymbol{\Sigma}$ is computed as $\boldsymbol{\Sigma} = \frac{1}{m}\mathbf{X}^{\rm T}\mathbf{X}$, where $\mathbf{X}^{\rm T}$ is the transpose of **X**. The symmetric $n \times n$ covariance matrix represents the correlations between variations in all $n(n+1)/2$ pairs of elements in the normalized maps. Eigendecomposition of $\boldsymbol{\Sigma}$ is performed, resulting in $n$ eigenvectors, $\mathbf{F}^i$, (each with $n$ elements), and corresponding eigenvalues, $\lambda^i$. The eigenvectors are the *principal components* of the data set, encoding the directions in $n$-space along which correlated variations in the data set are best described. The eigenvectors with the largest eigenvalues represent the majority of the variation in the data set. In the field of facial recognition, these eigenvectors are called eigenfaces. We will refer to our eigenvectors as eigenFACs.

These eigenFACs can be interpreted as maps indicating features which are consistently present in the data set. The 12 most significant eigenFACs are presented in Figure 2. The percentage of the variance in the data set described by eigenFAC $\mathbf{F}^i$ is given by $100\lambda^i / \sum_{j=1}^{n} \lambda^j$. Figure 2 (bottom) shows this contribution for the first 20 eigenFACs; the $\lambda^i = 1$ level is shown by the horizontal dashed line. The cumulative variance explained by the first 300 eigenFACs is indicated in Figure 2 (bottom). For reasons that will be discussed below, we also computed the eigenFACs for just the months June and July 2010–2012 ($m = 24251$, 93% success rate), presented in Figure 3. The eigenFACs calculated from the observations for all months are labeled $\mathbf{F}_A^i$ (Figure 2), those from summer months are labeled $\mathbf{F}_S^i$ (Figure 3).

The PCA technique decomposes the data set into as many eigenvectors as there are elements in each data vector, in our case $n = 960$. Which of these is significant in describing the data set? One of the aims of PCA is dimension-reduction: there are 960 elements in each FAC map, but can the gross structure of the data be described by just a few variables, and how many variables are necessary? The eigenvalue associated with each eigenvector is a measure of this significance. However, there is no hard-and-fast rule for determining the threshold for significance, though there are several commonly applied heuristics. Kaiser's criterion states that eigenvectors for which $\lambda^i \geq 1$ are significant [*Kaiser*, 1960]. In our case, this includes the first five or so eigenFACs. The Scree test [*Cattell*, 1966] orders the eigenvalues, as in Figures 2 and 3, and then fits a straight line through the lower eigenvalues and retains eigenvectors whose eigenvalues rise above this line. Application of the Scree test to our data indicates that the first 20 or so eigenvectors should be considered significant. On the other hand, another measure of significance is the number of eigenvectors necessary to explain the majority, say 95%, of the variance in the data set [*Hair et al.*, 1995]. In our case, approximately 70 eigenFACs are required to describe 80% of the variance and near 150 for 90% of the variance.

We first discuss the eigenFACs presented in Figure 2. The principal eigenFAC, $\mathbf{F}_A^1$, corresponding to approximately 25% of the variance in the data set, has the form of a concentric pair of upward/downward rings of Birkeland current, of opposite polarity in the dawn and dusk sectors, rotated such that the polarity changes along the 11/23 MLT meridians. This clearly represents the region 1 and 2 current system described by *Iijima and Potemra* [1976]. The rotation of the current rings is consistent with the rotation often seen in the ionospheric convection pattern [e.g., *Ruohoniemi and Greenwald*, 1996]. The magnitude of the contribution of this eigenFAC, $\alpha_A^1$, to a given current map, **J**, will represent the strength of magnetospheric convection. It might be expected that $\alpha_A^1$ will become large when the interplanetary magnetic field is negative, IMF $B_Z < 0$, and low-latitude magnetopause reconnection is ongoing. We note that $\mathbf{F}_A^1$ is consistent with the first principal component found by *Cousins et al.* [2015] in AMPERE data and by *Kim et al.* [2012] in their decomposition of polar cap ionospheric flow observations.

EigenFAC $\mathbf{F}_A^2$, representing approximately 10% of the variance, is a bipolar pair of latitudinally separated currents, strongest in the noon sector. The presence of $\mathbf{F}_A^2$, for which the coefficient $\alpha_A^2$ could be positive or negative, will modify the $\mathbf{F}^1_A$ pattern such that the R1 currents in the noon sector become upward or downward, with opposite polarity currents at higher latitude. This is consistent with currents expected to be associated with magnetic tension forces on newly reconnected field lines for nonzero IMF $B_Y$ [*Cowley et al.*, 1991]. $\mathbf{F}^2_A$ also contains a region of nightside R1 currents which possibly contribute to a dawn-dusk asymmetry in the convection flows on the nightside associated with IMF $B_Y$ (see, for instance, the discussion in *Milan* [2015]). Again, we note that the dayside portion of $\mathbf{F}^2_A$ is consistent with the second principal component found in the study of *Kim et al.* [2012]. It is also consistent with the third principle component found by [*Cousins et al.*, 2015]; their second principal component was associated with expansions and contractions of the current ovals, an effect that we have removed by preprocessing the current maps. $\mathbf{F}^3_A$ has a similar form to $\mathbf{F}^2_A$, though suggesting a more complicated, three-current structure on the dayside associated with IMF $B_Y$.

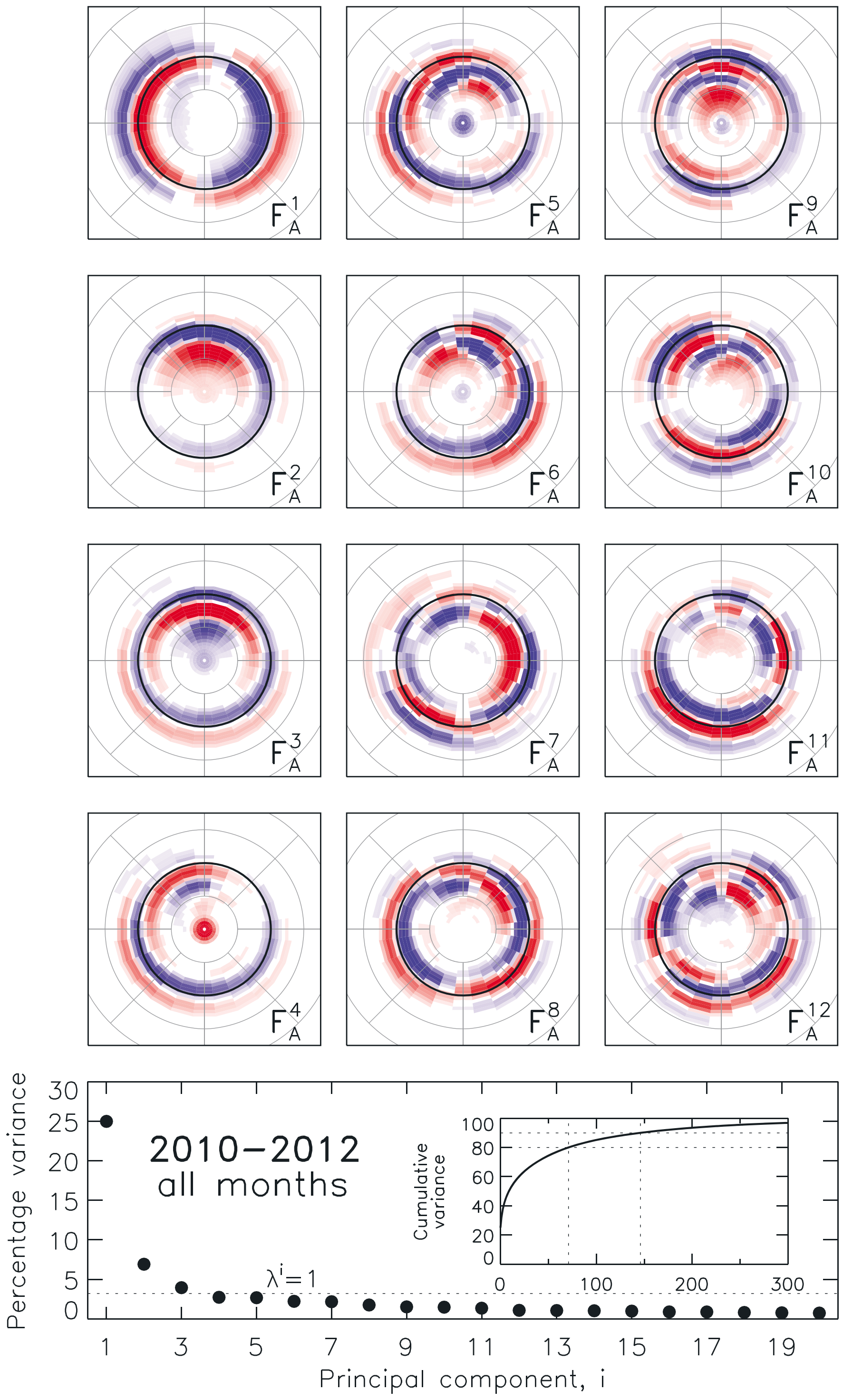

**Figure 2.** (top four rows) The first twelve eigenFACs, labeled $\mathbf{F}^i_A$, computed from the whole AMPERE data set from 2010 to 2012. Red and blue show FACs of opposite polarities but should not necessarily be identified as upward or downward. (bottom) The percentage of the variance in the total data set explained by each of the first 20 eigenFACs, proportional to the corresponding eigenvalue, $\lambda^i_A$. The horizontal dashed line shows the $\lambda^i_A = 1$ level. The inset shows the cumulative variance explained by the first 300 eigenFACs, with the 80% and 90% variance levels marked by horizontal dashed lines, and the corresponding eigenFACs by vertical dashed lines.

EigenFACs $\mathbf{F}^3_A$ and beyond contain significant currents on the nightside, in the main consisting of east-west aligned current sheets, which are most likely associated with substorm processes. The substorm current wedge is formed from upward and downward currents at the western and eastern edges of the substorm auroral bulge [*McPherron et al.*, 1973]. More than one eigenFAC contains upward/downward current pairs straddling the midnight meridian, e.g., $\mathbf{F}_A^5$, $\mathbf{F}_A^7$, and $\mathbf{F}_A^8$. As discussed above, these eigenFACs have low significance in comparison to $\mathbf{F}_A^1$ and $\mathbf{F}_A^2$, suggesting that there is not one eigenFAC that represents substorm current systems, but that these current systems are made from a variable superposition of a number of basis functions. As will be demonstrated below, although these eigenFACs have relatively low significance, they respond in a reproducible way to substorm onset.

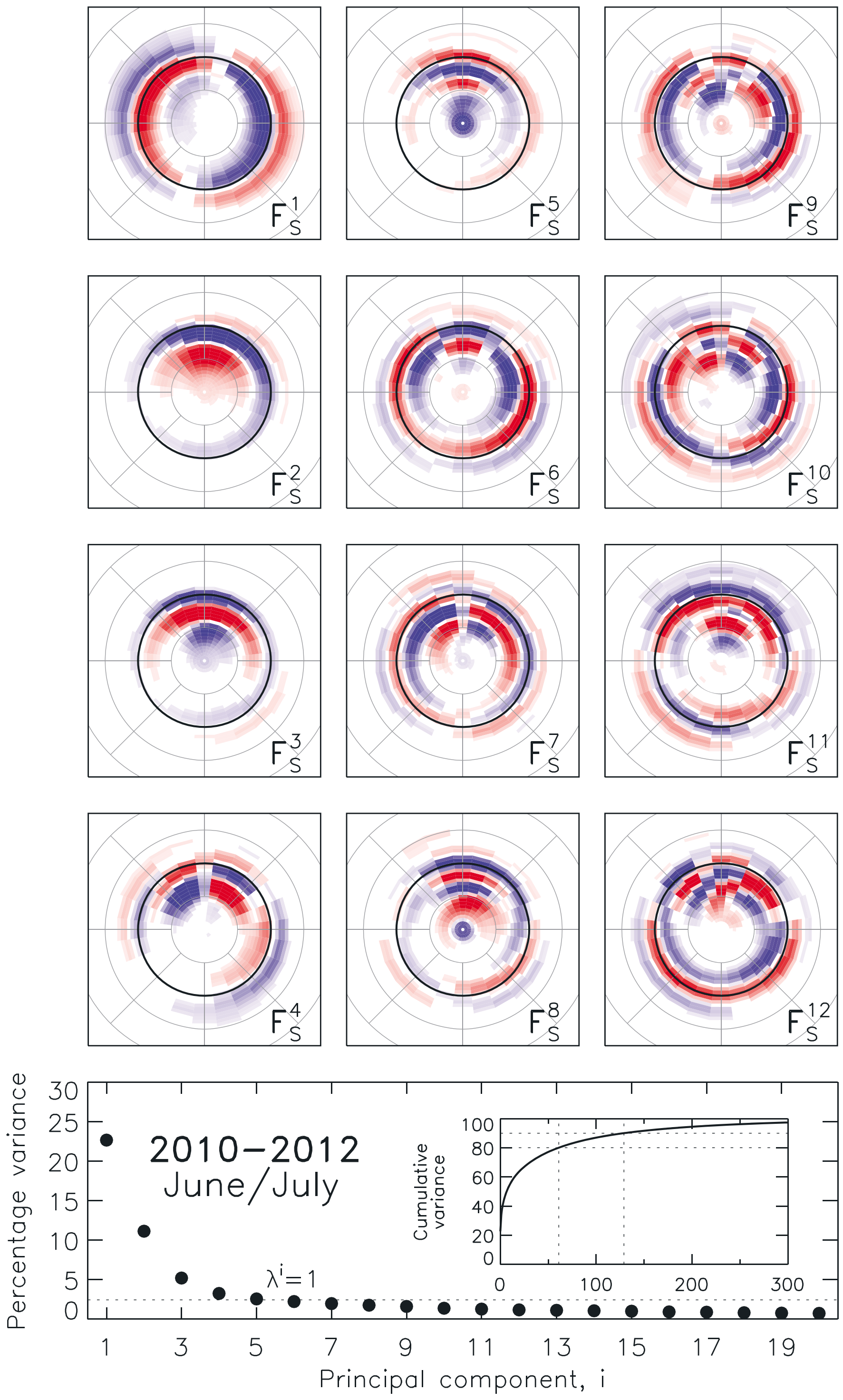

**Figure 3.** The same format as Figure 2 though for summer months June and July in 2010 to 2012, labeled $\mathbf{F}_S^i$.

Turning to Figure 3, the first three eigenFACs are very similar to their counterparts in Figure 2. However, eigenFAC $\mathbf{F}_S^4$ represents a quadrupolar region of currents near noon, consistent with reverse convection cells associated with lobe reconnection occurring for positive IMF $B_Z$. This eigenFAC also has currents in the dawn and dusk sectors of opposite polarity to the usual R1/R2 sense. This indicates that when the IMF is northward and lobe cells appear near noon, the R1/R2 currents at dawn and dusk are reduced. $\mathbf{F}_S^6$ in Figure 3 seems likely to be associated with IMF $B_Y$ changes in the location of the lobe reconnection site and distortions of the lobe cells [see, e.g., *Milan et al.*, 2000]. We note that although the R1/R2 current system of $\mathbf{F}_S^1$ is rotated from the noon-midnight meridian, most cusp-related features in Figure 3 are symmetric about the noon meridian. Other differences between the eigenFACs of Figures 2 and 3 will be discussed further below.

The eigenvectors produced by PCA form an orthonormal basis set, such that linear combinations of eigenFACs can be used to represent any current map. The contribution of a given eigenFAC, $\mathbf{F}^i$, to an individual map, $\mathbf{J}$, can be found from the inner product of $\mathbf{F}^i$ and $\mathbf{J}$: $\alpha^i = \sum_{j=1}^{n} F_j^i J_j$, where $J_j$ are the elements of $\mathbf{J}$ and $F_j^i$ are the elements of $\mathbf{F}^i$. Then

$$\mathbf{J} = \alpha^1\mathbf{F}^1 + \alpha^2\mathbf{F}^2 + \alpha^3\mathbf{F}^3 + \ldots \quad (1)$$

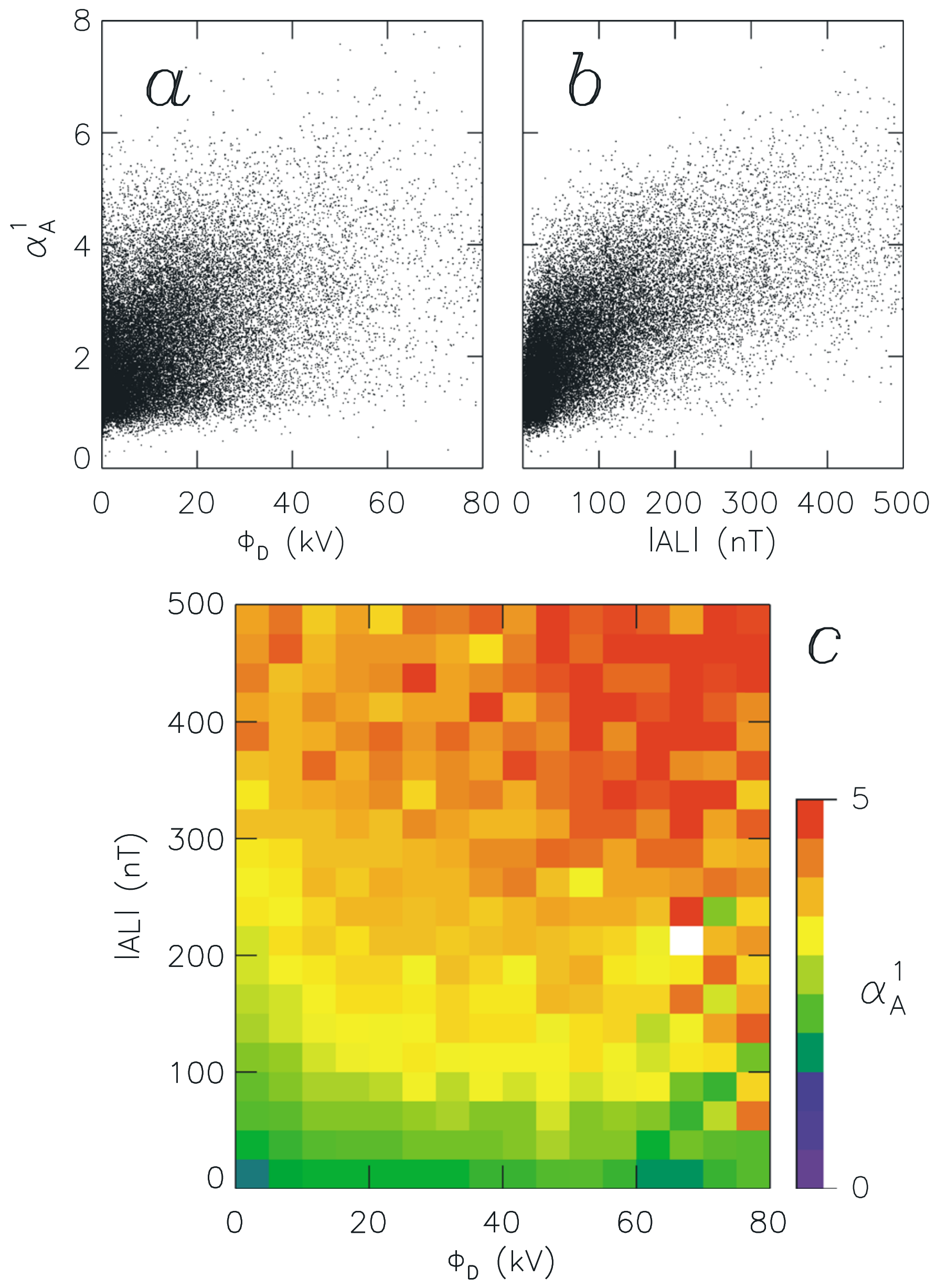

**Figure 4.** (a) The projection, $\alpha_A^1$, of eigenFAC $\mathbf{F}_A^1$ on each current map $\mathbf{J}$ from 2010 as a function of dayside reconnection rate, $\Phi_D$, predicted from upstream solar wind conditions. (b) $\alpha_A^1$ as a function of $|AL|$, which we use as a proxy for the nightside reconnection rate, $\Phi_N$. (c) The mean value of $\alpha_A^1$ as a function of $\Phi_D$ and $|AL|$.

That is, $\alpha^i$ is the projection of $\mathbf{J}$ along a particular basis vector $\mathbf{F}^i$. This procedure can be used to find the contribution of each eigenFAC to the observed currents in a particular map, and how these change with time or geomagnetic activity.

As discussed above, $\mathbf{F}_A^1$ is expected to be associated with large-scale convection, so $\alpha_A^1$ should be a measure of convection strength, and hence dependent on the activity of the substorm cycle. Figure 4a shows the contribution, $\alpha_A^1$, of $\mathbf{F}_A^1$ to each current map $\mathbf{J}$ from 2010, as a function of dayside reconnection rate $\Phi_D$, determined using the parameterization of *Milan et al.* [2012] and 1 min OMNI solar wind data [*King and Papitashvili*, 2005]. Figure 4b shows $\alpha_A^1$ as a function of the absolute value of the *AL* electrojet index, which we use as a proxy for the nightside reconnection rate $\Phi_N$ [see also *Holzer et al.*, 1986 and *Coxon et al.*, 2014a, 2014b]. $\alpha_A^1$ increases with both $\Phi_D$ and $\Phi_N$, though the dependence is rather weak for $\Phi_D$. The expanding/contracting polar cap model [*Cowley and Lockwood*, 1992; *Milan et al.*, 2007] suggests that the cross-polar cap potential should be dependent on both dayside and nightside reconnection. Figure 4c shows the mean value of $\alpha_A^1$ as a function of both $\Phi_D$ and $|AL|$, confirming that $\alpha_A^1$ increases for both, consistent with the findings of *Coxon et al.* [2014a].

We expect $\alpha_A^2$ to be associated with tension forces due to the $B_Y$ component of the IMF. Figure 5 presents this dependence for 2010. In this case we find a strong seasonal variation in $\alpha_A^2$, so we have separated the data by month. At all times there is a clear correlation, but the magnitude of $\alpha_A^2$ is larger in summer months than in winter months. This suggests that the conductance of the ionosphere at the footprint of newly reconnected field lines significantly modulates the magnitude of the currents associated with tension forces. It is this seasonal dependence that motivated us to produce the eigenFACs for June and July shown in Figure 3. Comparing the $\mathbf{F}_A^2$ to $\mathbf{F}_A^5$ (Figure 2) with $\mathbf{F}_S^2$ to $\mathbf{F}_S^5$ (Figure 3), it is clear that noon sector currents are more pronounced with respect to nightside currents in summer months (Figure 3) than in the eigenFACs computed from the whole data set (Figure 2).

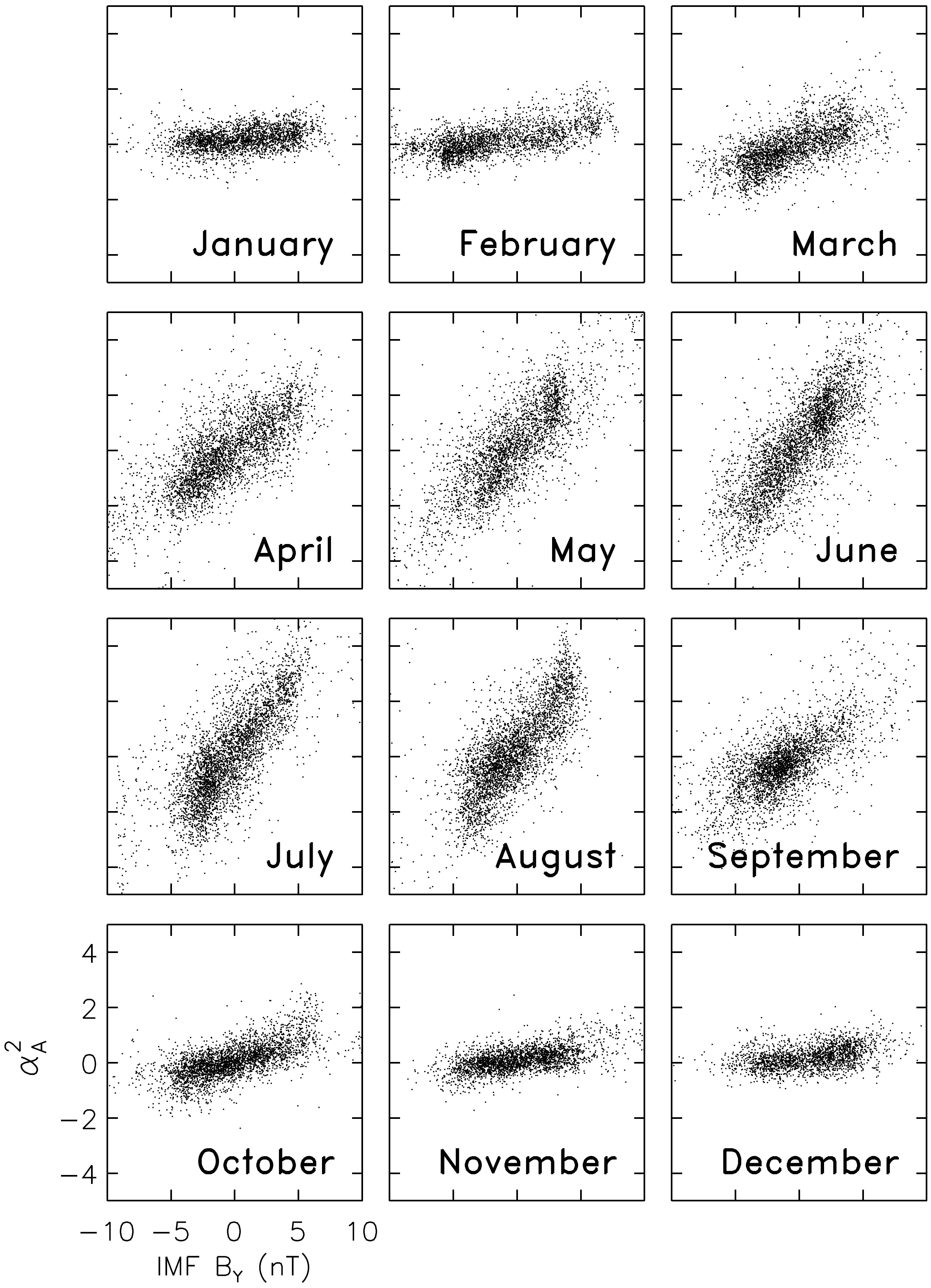

**Figure 5.** The projection, $\alpha_A^2$, of eigenFAC $\mathbf{F}_A^2$ on each current map $\mathbf{J}$ as a function IMF $B_Y$ for each month of 2010.

EigenFACs with significant nightside currents are expected to respond to the occurrence of substorms. As already discussed, it seems that no one eigenFAC captures the substorm dynamic, otherwise such an eigenFAC would be expected to have a relatively large eigenvalue. Rather, although Kaiser's criterion suggests that eigenFACs beyond $\mathbf{F}_A^5$ have low significance, they contribute in a collective manner to describe the variance observed in different substorms. Here we test if the eigenFACs show some consistency in their response to substorm onset. We do this by performing a superposed epoch analysis of the $\alpha_A^i$s, in which we use approximately 3000 substorm expansion phase onsets as determined by SuperMAG [*Newell and Gjerloev*, 2011] as the zero epoch. The results are presented in Figure 6.

Figure 6 (first panel) shows the variation of $\Lambda_{\mathrm{fit}}$, the radius of the current ovals used in the normalization procedure. The ovals expand to lower latitude prior to onset, continue to expand until 20 min after onset, before contracting to higher latitudes again, consistent with the ECPC. Figure 6 (second to sixth panels) presents the $\alpha_A^i$s which show the largest variation; vertical bars indicate the standard error on the mean. $\alpha_A^1$ shows a small increase prior to onset (convection associated with dayside reconnection and the growth phase) and a large increase after onset (associated with convection excited by nightside reconnection). Both $\Lambda_{\rm fit}$ and $\alpha_A^1$ results are consistent with the findings of *Coxon et al.* [2014b].

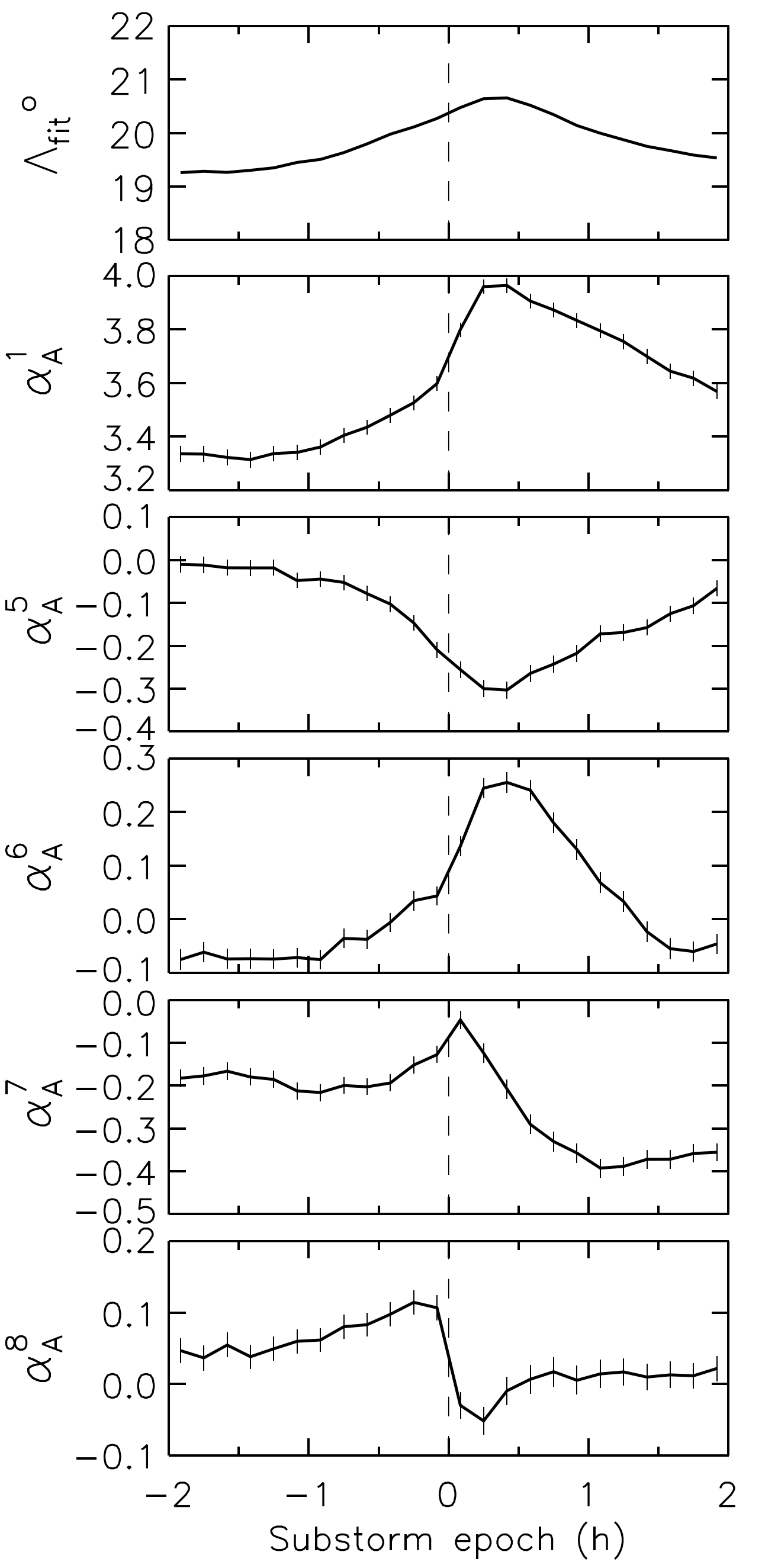

**Figure 6.** (first panel) Superposed epoch analysis of $\Lambda_{\rm fit}$ and the (second to sixth panels) projections, $\alpha_A^i$, of several major eigenFACs, with substorm onset determined by SuperMAG as the zero epoch, for 2010 to 2012. Vertical bars indicate the standard error on the mean.

Projections $\alpha_A^2$ to $\alpha_A^4$ are not shown in Figure 6 as their variation with substorms is small. This is expected as they largely describe the effects of the dayside tension forces associated with IMF $B_Y$, which should not vary consistently with substorm cycle.

The other $\alpha_A^i$s all show variations associated with the growth and/or expansion phases. For instance, $\alpha_A^5$ is close to zero before the growth phase but becomes increasingly negative during the expansion phase. The associated eigenFAC, $\mathbf{F}_A^5$, has currents on the nightside that are of opposite sense to the R1/R2 current system represented by $\mathbf{F}_A^1$, suggesting that $\mathbf{F}_A^5$ when subtracted from $\mathbf{F}_A^1$ represents an enhancement in nightside R1/R2 currents during substorms. This is consistent with an enhancement of nightside convection associated with magnetotail reconnection and the enhancements in currents due to conductance changes in the nightside auroral zone. Projections $\alpha_A^6$ and beyond also show significant substorm-correlated changes. The corresponding eigenFACs have increasingly structured currents on the nightside which will contribute to the form of the substorm current wedge. Figure 6 suggests that the substorm FAC structure evolves rapidly after substorm onset, especially with respect to the 10 min cadence of the AMPERE observations: for this reason, a single eigenFAC cannot represent the substorm current wedge at all substorm phases. Moreover, despite the low significance attributed to principal components beyond $\mathbf{F}_A^5$, their consistent response to substorm onset suggests that they do represent repeatable structures within the nightside current systems. In part, their low significance will arise as a consequence of the small proportion of time they are active, whereas the R1/R2 currents and dayside tension forces ($\mathbf{F}_A^1$ and $\mathbf{F}_A^2$) will be present much of the time.

The ECPC predicts that magnetospheric convection is associated with the action of both dayside and nightside reconnection (in fact, that the cross-polar cap potential is the average of $\Phi_D$ and $\Phi_N$) [*Lockwood*, 1991]. However, two of our observations seem to somewhat contradict this: in Figure 4a the dependence of $\alpha_A^1$ on $\Phi_D$ is weaker than for $|AL|$, and in Figure 6 the increase in $\alpha_A^1$ is much more significant after substorm onset than during the growth phase. We interpret this as the control of current magnitudes by the nightside ionospheric conductance, which will be enhanced during substorms, but not necessarily during the growth phase.

Finally, we discuss some similarities and differences between the eigenFACs calculated for the whole period 2010–2012, $\mathbf{F}_A^i$ (Figure 2), and for the months June and July alone, $\mathbf{F}_S^i$ (Figure 3). $\mathbf{F}^1$, $\mathbf{F}^2$, and $\mathbf{F}^3$ are very similar in both sets of eigenFACs, though we note that the variance represented by $\mathbf{F}_S^2$ and $\mathbf{F}_S^3$ is relatively larger than $\mathbf{F}_A^2$ and $\mathbf{F}_A^3$, recognizing the enhanced cusp currents when the dayside ionospheric conductance is high. The nightside currents in $\mathbf{F}_S^3$ are weaker than in $\mathbf{F}_A^3$, again representing the dominance and consistency of noon currents in summer.

It becomes more difficult to see direct relationships between other pairs of eigenFACs in Figures 2 and 3 as the eigenvalues decrease. Indeed, many of these eigenFACs appear to share common features of dayside/nightside currents, but the contributions differ between the two data sets. This reinforces the point discussed above that substorm FACs do not have one dominant mode or spatial pattern but change in a subtle but complicated way from substorm to substorm. Hence, no single mode of response is identified by the PCA technique. Despite this, it is anticipated that a more in-depth study of these eigenFACs could lead to a greater understanding of substorm processes.

There are two additional factors to consider in future studies: the dependence of substorm behavior on the latitude of substorm onset and the IMF $B_Y$ dependence of magnetotail dynamics. *Milan et al.* [2009] demonstrated that the auroral signature of substorms is enhanced when the substorms occur on an expanded auroral oval (greater proportion of open magnetic flux at onset), and *Grocott et al.* [2009] showed corresponding changes in the ionospheric convection response to substorms. A similar dependence may be expected to occur in the contributions of different eigenFACs during substorms of differing onset latitudes. It is also known that the local time of substorm onset can be influenced by the $B_Y$ component of the IMF some time prior to onset [e.g., *Østgaard et al.*, 2005; *Milan et al.*, 2010], and this might be reflected in the eigenFACs.

## 3. Conclusions

We have presented a principal component analysis (PCA) of Birkeland current measurements from the AMPERE experiment. The technique reveals that the region 1/2 current system first identified by *Iijima and Potemra* [1976] is the most reproducible feature of the data set. Dayside currents associated with magnetic tension forces on newly reconnected field lines are the second most consistent feature. The strength of these cusp currents is significantly modulated by season, being strongest in summer months when the ionospheric conductance at the footprint of the cusp region is greatest. This will introduce a significant interhemispheric asymmetry in the currents flowing on the dayside magnetopause. Cusp currents associated with lobe reconnection and reverse convection are also apparent in the decomposition during summer months. In contrast, nightside current patterns are complicated and do not have a consistent response to substorm onset—there is no easily identifiable substorm current wedge—though the magnitude of the R1/R2 currents is enhanced.

**Acknowledgments**
S.E.M. and J.A.C. were supported by the Science and Technology Facilities Council (STFC), UK, grant ST/K001000/1. The work at the Birkeland Centre for Space Centre, University of Bergen, Norway, was supported by the Research Council of Norway/CoE under contract 223252/F50. We thank the AMPERE team and the AMPERE Science Center for providing the Iridium-derived data products; AMPERE products are available at http://ampere.jhuapl.edu. The OMNI data, including solar wind parameters and geomagnetic indices, were obtained from the GSFC/SPDF OMNIWeb interface at http://omniweb.gsfc.nasa.gov. The SuperMAG substorm list was downloaded from http://supermag.jhuapl.edu. For the ground magnetometer data from which this list was derived, we gratefully acknowledge the following: Intermagnet; USGS, Jeffrey J. Love; CARISMA, PI Ian Mann; CANMOS; the S-RAMP Database, PI K. Yumoto and K. Shiokawa; The SPIDR database; AARI, PI Oleg Troshichev; the MACCS program, PI M. Engebretson, Geomagnetism Unit of the Geological Survey of Canada; GIMA; MEASURE, UCLA IGPP and Florida Institute of Technology; SAMBA, PI Eftyhia Zesta; 210 Chain, PI K. Yumoto; SAMNET, PI Farideh Honary; the institutes who maintain the IMAGE magnetometer array, PI Eija Tanskanen; PENGUIN; AUTUMN, PI Martin Connors; DTU Space, PI Juergen Matzka; South Pole and McMurdo Magnetometer, PIs Louis J. Lanzarotti and Alan T. Weatherwax; ICESTAR; RAPIDMAG; PENGUIn; British Artarctic Survey; McMac, PI Peter Chi; BGS, PI Susan Macmillan; Pushkov Institute of Terrestrial Magnetism, Ionosphere and Radio Wave Propagation (IZMIRAN); GFZ, PI Juergen Matzka; MFGI, PI B. Heilig; IGFPAS, PI J. Reda; University of L'Aquila, PI M. Vellante; and SuperMAG, PI Jesper W. Gjerloev. S.E.M. would like to thank Colin Waters, University of Newcastle, Australia, for helpful discussions regarding AMPERE data and principal component analysis.

## References

Anderson, B. J., K. Takahashi, and B. A. Toth (2000), Sensing global Birkeland currents with Iridium® engineering magnetometer data, *Geophys. Res. Lett.*, *27*, 4045–4048, doi:10.1029/2000GL000094.

Anderson, B. J., K. Takahashi, T. Kamei, C. L. Waters, and B. A. Toth (2002), Birkeland current system key parameters derived from Iridium observations: Method and initial validation results, *J. Geophys. Res.*, *107*, 1079, doi:10.1029/2001JA000080.

Anderson, B. J., H. Korth, C. L. Waters, D. L. Green, and P. Stauning (2008), Statistical Birkeland current distributions from magnetic field observations by the Iridium constellation, *Ann. Geophys.*, *26*, 671–687, doi:10.5194/angeo-26-671-2008.

Anderson, B. J., H. Korth, C. L. Waters, D. L. Green, V. G. Merkin, R. J. Barnes, and L. P. Dyrud (2014), Development of large-scale Birkeland currents determined from the Active Magnetosphere and Planetary Electrodynamics Response Experiment, *Geophys. Res. Lett.*, *41*, 3017–3025, doi:10.1002/2014GL059941.

Cattell, R. B. (1966), The Scree test for the number of factors, *Multivariate Behavioral Res.*, *1*, 245–276.

Clausen, L. B. N., J. B. H. Baker, J. M. Ruohoniemi, S. E. Milan, and B. J. Anderson (2012), Dynamics of the region 1 Birkeland current oval derived from the Active Magnetosphere and Planetary Electrodynamics Response Experiment (AMPERE), *J. Geophys. Res.*, *117*, A06233, doi:10.1029/2012JA017666.

Clausen, L. B. N., J. B. H. Baker, J. M. Ruohoniemi, S. E. Milan, J. C. Coxon, S. Wing, S. Ohtani, and B. J. Anderson (2013a), Temporal and spatial dynamics of the region 1 and 2 Birkeland currents during substorms, *J. Geophys. Res. Space Physics*, *118*, 3007–3016, doi:10.1002/jgra.50288.

Clausen, L. B. N., S. E. Milan, J. B. H. Baker, J. M. Ruohoniemi, K.-H. Glassmeier, J. C. Coxon, and B. J. Anderson (2013b), On the influence of open magnetic flux on substorm intensity: Ground- and space-based observations, *J. Geophys. Res. Space Physics*, *118*, 2958–2969, doi:10.1002/jgra.50308.

Cousins, E. D. P., T. Matsuo, A. D. Richmond, and B. J. Anderson (2015), Dominant modes of variability in large-scale Birkeland currents, *J. Geophys. Res. Space Physics*, *120*, 6722–6735, doi:10.1002/2014JA020462.

Cowley, S. W. H., and M. Lockwood (1992), Excitation and decay of solar wind-driven flows in the magnetosphere-ionosphere system, *Ann. Geophys.*, *10*, 103–115.

Cowley, S. W. H., J. P. Morelli, and M. Lockwood (1991), Dependence of convective flows and particle precipitation in the high-latitude dayside ionosphere on the $X$ and $Y$ components of the interplanetary magnetic field, *J. Geophys. Res.*, *96*, 5557–5564.

Coxon, J. C., S. E. Milan, L. B. N. Clausen, B. J. Anderson, and H. Korth (2014a), The magnitudes of the regions 1 and 2 Birkeland currents observed by AMPERE and their role in solar wind-magnetosphere-ionosphere coupling, *J. Geophys. Res. Space Physics*, *119*, 9804–9815, doi:10.1002/2014JA020138.

Coxon, J. C., S. E. Milan, L. B. N. Clausen, B. J. Anderson, and H. Korth (2014b), A superposed epoch analysis of the Region 1 and Region 2 Birkeland currents observed by AMPERE during substorms, *J. Geophys. Res. Space Physics*, *119*, 9834–9846, doi:10.1002/2014JA020500.

Dungey, J. W. (1961), Interplanetary magnetic fields and the auroral zones, *Phys. Rev. Lett.*, *6*, 47–48.

Grocott, A., J. A. Wild, S. E. Milan, and T. K. Yeoman (2009), Superposed epoch analysis of the ionospheric convection evolution during substorms: Onset latitude dependence, *Ann. Geophys.*, *27*, 591–600.

Hair, J., R. E. Anderson, R. L. Tatham, and W. C. Black (1995), *Multivariate Data Analysis*, 4th ed., Prentice-Hall, N. J.
He, M., J. Vogt, H. Lühr, E. Sorbalo, A. Blagau, G. Le, and G. Lu (2012), A high-resolution Model of Field-Aligned Currents Through Empirical orthogonal functions analysis (MFACE), *Geophys. Res. Lett.*, *39*, L18105, doi:10.1029/2012GL053168.
Holzer, R. E., R. L. McPherron, and D. A. Hardy (1986), A quantitative empirical model of the magnetospheric flux transfer process, *J. Geophys. Res.*, *91*(A3), 3287–3293, doi:10.1029/JA091iA03p03287.
Iijima, T., and T. A. Potemra (1976), The amplitude distribution of field-aligned currents at northern high latitudes observed by Triad, *J. Geophys. Res.*, *81*, 2165–2174, doi:10.1029/JA081i013p02165.
Jolliffe, I. T. (2002), *Principal Component Analysis*, 2nd ed., Springer, New York.
Kaiser, H. F. (1960), The application of electronic computers to factor analysis, *Educ. Psychol. Meas.*, *20*, 141–151, doi:10.1177/001316446002000116.
Kim, H.-J., L. R. Lyons, J. M. Ruohoniemi, N. A. Frissel, and J. B. Baker (2012), Principal component analysis of polar cap convection, *Geophys. Res. Lett.*, *39*, L11105, doi:10.1029/2012GL052083.
King, J. H., and N. E. Papitashvili (2005), Solar wind spatial scales in and comparisons of hourly Wind and ACE plasma and magnetic field data, *J. Geophys. Res.*, *110*, A02209, doi:10.1029/2004JA010804.
Lockwood, M. (1991), On flow reversal boundaries and transpolar voltage in average models of high latitude convection, *Planet. Space Sci.*, *3*, 397–409.
Lockwood, M., and S. W. H. Cowley (1992), Ionospheric convection and the substorm cycle, in *Substorms 1, Proceedings of the First International Conference on Substorms, ICS-1*, edited by C. Mattock, pp. 99–109, ESA-SP-335 European Space Agency Publ., Nordvijk, Netherlands.
McPherron, R. L., C. T. Russell, and M. P. Aubry (1973), Satellite studies of magnetospheric substorms on August 15, 1968. 9. Phenomenological model for substorms, *J. Geophys. Res.*, *78*, 3131–3149.
Milan, S. E. (2009), Both solar wind-magnetosphere coupling and ring current intensity control of the size of the auroral oval, *Geophys. Res. Lett.*, *36*, L18101, doi:10.1029/2009GL039997.
Milan, S. E. (2015), Sun et Lumière: Solar wind-magnetosphere coupling as deduced from ionospheric flows and polar auroras, in *Magnetospheric Plasma Physics: The Impact of Jim Dungey's Research, Astrophys. and Space Sci. Proc.*, vol. 41, edited by D. Southwood et al., Springer, Switzerland, doi:10.1007/978-3-319-18359-6_2.
Milan, S. E., M. Lester, S. W. H. Cowley, and M. Brittnacher (2000), Dayside convection and auroral morphology during an interval of northward interplanetary magnetic field, *Ann. Geophys.*, *18*, 436–444.
Milan, S. E., M. Lester, S. W. H. Cowley, K. Oksavik, M. Brittnacher, R. A. Greenwald, G. Sofko, and J.-P. Villain (2003), Variations in polar cap area during two substorm cycles, *Ann. Geophys.*, *21*, 1121–1140.
Milan, S. E., G. Provan, and B. Hubert (2007), Magnetic flux transport in the Dungey cycle: A survey of dayside and nightside reconnection rates, *J. Geophys. Res.*, *112*, A01209, doi:10.1029/2006JA011642.
Milan, S. E., A. Grocott, C. Forsyth, S. M. Imber, P. D. Boakes, and B. Hubert (2009), A superposed epoch analysis of auroral evolution during substorm growth, onset and recovery: Open magnetic flux control of substorm intensity, *Ann. Geophys.*, *27*, 659–668.
Milan, S. E., A. Grocott, and B. Hubert (2010), A superposed epoch analysis of auroral evolution during substorms: Local time of onset region, *J. Geophys. Res.*, *115*, A00I04, doi:10.1029/2010JA015663.
Milan, S. E., J. S. Gosling, and B. Hubert (2012), Relationship between interplanetary parameters and the magnetopause reconnection rate quantified from observations of the expanding polar cap, *J. Geophys. Res.*, *117*, A03226, doi:10.1029/2011JA017082.
Murphy, K. R., I. R. Mann, I. J. Rae, C. L. Waters, H. U. Frey, A. Kale, H. J. Singer, B. J. Anderson, and H. Korth (2013), The detailed spatial structure of field-aligned currents comprising the substorm current wedge, *J. Geophys. Res. Space Physics*, *118*, 7714–7727, doi:10.1002/2013JA018979.
Natali, M. P., and A. Meza (2010), Annual and semiannual VTEC effects at low solar activity based on GPS observations at different geomagnetic latitudes, *J. Geophys. Res.*, *115*, D18106, doi:10.1029/2010JD014267.
Newell, P. T., and J. W. Gjerloev (2011), Evaluation of SuperMAG auroral electrojet indices as indicators of substorms and auroral power, *J. Geophys. Res.*, *116*, A12211, doi:10.1029/2011JA016779.
Ohme, F., A. B. Nielsen, D. Keppel, and A. Lundgren (2013), Statistical and systematic errors for gravitational-wave inspiral signals: A principal component analysis, *Phys. Rev. D*, *88*, 42002, doi:10.1103/PhysRevD.88.042002.
Østgaard, N., N. A. Tsyganenko, S. B. Mende, H. U. Frey, T. J. Immel, M. Fillingim, L. A. Frank, and J. B. Sigwarth (2005), Observations and model predictions of substorm auroral asymmetries in the conjugate hemispheres, *Geophys. Res. Lett.*, *32*, L05111, doi:10.1029/2004GL022166.
Ruohoniemi, J. M., and R. A. Greenwald (1996), Statistical patterns of high-latitude convection obtained from Goose Bay HF radar observations, *J. Geophys. Res.*, *101*, 21,743–21,763.
Sergeev, V. A., et al. (2014), Event study combining magnetospheric and ionospheric perspectives of the substorm current wedge modeling and dynamics, *J. Geophys. Res. Space Physics*, *119*, 9714–9728, doi:10.1002/2014JA020522.
Sirovich, L., and M. Kirby (1987), Low-dimensional procedure for the characterization of human faces, *J. Opt. Soc. Am. A*, *4*, 519–524, doi:10.1364/JOSAA.4.000519.
Siscoe, G. L., and T. S. Huang (1985), Polar cap inflation and deflation, *J. Geophys. Res.*, *90*, 543–547.
Turk, M., and A. Pentland (1991), Eigenfaces for recognition, *J. Cognitive Neurosci.*, *3*, 71–86, doi:10.1162/jocn.1991.3.1.71.
Waters, C. L., B. J. Anderson, and K. Liou (2001), Estimation of global field aligned currents using the Iridium®; System magnetometer data, *Geophys. Res. Lett.*, *28*, 2165–2168.
Wilder, F. D., S. Eriksson, H. Korth, J. B. H. Baker, M. R. Hairston, C. Heinselman, and B. J. Anderson (2013), Field-aligned current reconfiguration and magnetospheric response to an impulse in the interplanetary magnetic field $B_Y$ component, *Geophys. Res. Lett.*, *40*, 2489–2494, doi:10.1002/grl.50505.